\documentclass[prx,aps,groupedaddress,twocolumn]{revtex4-1}
\usepackage{graphicx}
\usepackage{dcolumn}
\usepackage{bm}
\usepackage{color}

\usepackage{epstopdf}

\usepackage{amsmath, amssymb, graphics, setspace}

\newcommand{\mathsym}[1]{{}}
\newcommand{\unicode}[1]{{}}

\usepackage{titlesec}
\usepackage{bm}
\usepackage{comment}
\usepackage{verbatim}

\usepackage{graphicx}
\usepackage{subfigure}
\usepackage{tabularx}

\usepackage{color}
\usepackage[colorlinks,bookmarks=false,citecolor=blue,linkcolor=red,urlcolor=blue]{hyperref}

\definecolor{darkred}{rgb}{0.7,0.0,0.0}

\definecolor{darkblue}{rgb}{0,0.02,0.45}

\definecolor{darkgreen}{rgb}{0.02,0.45,0.0}

\definecolor{violet}{rgb}{0.8,0.2,0.6}

\def\be{\begin{equation}}
\def\ee{\end{equation}}
\def\bea{\begin{eqnarray}}
\def\eea{\end{eqnarray}}

\def\vec{\mathbf}

\def\mc{\mathcal}

\usepackage{times}

\begin{document}
\title{Magnetic field-induced evolution of intertwined orders in the  Kitaev magnet $\beta$-Li$_2$IrO$_3$}

\author{Ioannis Rousochatzakis}
\affiliation{School of Physics and Astronomy, University of Minnesota, Minneapolis,
MN 55116, USA}

\author{Natalia  B. Perkins}
\affiliation{School of Physics and Astronomy, University of Minnesota, Minneapolis,
MN 55116, USA}

\begin{abstract}
Recent scattering experiments in the 3D Kitaev magnet $\beta$-Li$_2$IrO$_3$ have shown that a relatively weak magnetic field along the crystallographic ${\bf b}$-axis drives the system from its incommensurate counter-rotating order to a correlated paramagnet, with a significant uniform `zigzag' component superimposing the magnetization along the field. 
Here it is shown that the zigzag order is not emerging from its linear coupling to the field (via a staggered, off-diagonal element of the ${\bf g}$-tensor), but from its intertwining with the incommensurate order and the longitudinal magnetization. 
The emerging picture explains all qualitative experimental findings at zero and finite fields, including the rapid decline of the incommensurate order with field and the so-called intensity sum rule. 
The latter are shown to be independent signatures of the smallness of the Heisenberg exchange $J$, compared to the Kitaev coupling $K$ and the off-diagonal  anisotropy $\Gamma$. 
Remarkably, in the regime of interest, the field $H^\ast$ at which the incommensurate component vanishes, depends essentially only on $J$, which allows to extract an estimate of $J\!\simeq\!4K$ from reported measurements of $H^\ast$.
We also comment on recent experiments in pressurized $\beta$-Li$_2$IrO$_3$ and conclude that $J$ decreases with pressure.
\end{abstract}
\maketitle

\section{Introduction}
\vspace*{-0.3cm}
The realization~\cite{Jackeli2009,Jackeli2010} that certain correlated materials based on 4d and 5d transition metals, like Ir$^{4+}$ or Ru$^{3+}$, host the key microscopic ingredients for the so-called Kitaev spin liquid~\cite{Kitaev2006} has spurred tremendous experimental and theoretical interest in the last decade~\cite{BookCao,Balents2014,Rau2016,Trebst2017,Hermanns2017,Winter2017}. 
A recurring theme in this research is that the predicted quantum spin liquids~\cite{Kitaev2006,Mandal2009,Kimchi2014,Hermanns2016} are fragile against various realistic perturbations~\cite{Jackeli2010,Schaffer2012,Jackeli2013,Lee2014,Katukuri2014,Katukuri2015,IoannisK1K2,Satoshi2016} and are preempted by magnetic order at low enough temperatures~\cite{
Singh2010, Singh2012,Liu2011,
Sears2015,Johnson2015,
Williams2016,
%
Biffin2014b, 
Modic2014,Biffin2014a,Takayama2015}. 
Nevertheless, there is overwhelming evidence that external perturbations, such as magnetic field~\cite{
Baek2017,Yadav2016,Sears2017,Zheng2017,Hentrich2017,Kataev2017,
Ruiz2017,Tsirlin2018,
Modic2017,
Chern2017,Janssen2016,Vojta2017,Winter2018
}, chemical substitution~\cite{Takagi2018}, or pressure~\cite{
Takayama2015,Veiga2017,Tsirlin2018,
Breznay2017,
Bastien2018,Biesner2018,
KimKimKee2016
}, can drive these materials to various types of correlated phases, including spin liquids. 
To go forward, it is therefore crucial to map out the most relevant instabilities, and identify their distinctive experimental fingerprints.

In this vein, we study the enigmatic magnetic-field induced instability reported recently by Ruiz {\it et al}~\cite{Ruiz2017} in the 3D hyperhoneycomb iridate $\beta$-Li$_2$IrO$_3$~\cite{Takayama2015,Biffin2014a}. 
At zero field, this magnet was known~\cite{Biffin2014a} to develop (below $T_N\!=\!37$ K) a counter-rotating incommensurate (IC) modulation, similar with those in the 3D stripy-honeycomb $\gamma$-Li$_2$IrO$_3$~\cite{Biffin2014b,Modic2014} and the layered honeycomb $\alpha$-Li$_2$IrO$_3$~\cite{Williams2016}.
The new findings from the magnetic resonant X-ray scattering data under a finite field are the following~\cite{Ruiz2017}:
(i) The IC order of $\beta$-Li$_2$IrO$_3$ is very fragile against a magnetic field along the crystallographic ${\bf b}$-axis, and disappears completely at a characteristic field $H^\ast$.  
(ii) The system develops a significant uniform `zigzag' component along ${\bf a}$ (superimposing the magnetization along ${\bf b}$), similar to the zigzag order of Na$_2$IrO$_3$~\cite{Singh2010,Singh2012,Liu2011,Choi2012,Ye2012,Chun2015} and $\alpha$-RuCl$_3$~\cite{Plumb2014,Sears2015,Kubota2015,Majumder2015,Banerjee2016, Johnson2015}.
(iii) The zigzag component grows linearly with field until it shows a kink at $H^\ast$, but is otherwise undetectable at zero field, consistent with the experiments of Biffin {\it et al}~\cite{Biffin2014a}. 
(iv) Quite surprisingly, the sum of the intensities of the Bragg peaks associated with the IC and the uniform components (rescaled by some factor) remains constant up to a field slightly larger than $H^\ast$.

One way to rationalize the appearance of the zigzag component is to build on the insight that a field along ${\bf b}$ couples linearly not only to the uniform magnetization along ${\bf b}$, but also to the zigzag component along ${\bf a}$, by virtue of the off-diagonal element $g_{ab}$ of the ${\bf g}$-tensor.~\cite{Ruiz2017}  This would explain the growth of the zigzag component (besides the magnetization along ${\bf b}$) at the expense of the IC order. However, this picture of a field-induced zigzag order cannot readily explain the significant zigzag amplitude at $H^\ast$, the intensity sum rule (point (iv) above), as well as the $T$-dependence of the intensity of the zigzag component, which is very similar to that of an order parameter~\cite{Ruiz2017}.

The results presented below reveal that a more consistent scenario is that the IC counter-rotating component, the zigzag component along ${\bf a}$ and the magnetization along ${\bf b}$ are intertwined components of the same order.  
It is shown in particular, that the significant growth of the zigzag component with field occurs even in the absence of the off-diagonal element $g_{ab}$. This demonstrates that the zigzag component does not originate in its linear coupling to the field, but rather in an intrinsic coupling with the IC order and the magnetization along ${\bf b}$. 
In fact, as shown in Ref.~[\onlinecite{Sam2018}], both the zigzag component and the magnetization along ${\bf b}$ are already present at zero field, albeit with an amplitude that is too weak to be detected. 

The emerging picture explains all the qualitative experimental  results at both zero~\cite{Biffin2014a} and finite fields~\cite{Ruiz2017}. First, the weak zigzag amplitude at zero-field, the rapid decline of the IC order with field, and the intensity sum rule are all facets of the same fact. Namely, that the Heisenberg coupling $J$ is much weaker than both the Kitaev interaction $K$ and the off-diagonal exchange anisotropy $\Gamma$. Second, the characteristic field $H^\ast$ is essentially independent of the dominant interactions $K$ and $\Gamma$ and scales linearly with $J$. This allows to deduce an estimate of $J\!\simeq\!4$\,K from reported data of $H^\ast$. Furthermore, a comparison with recent experiments under pressure~\cite{Tsirlin2018} suggests that $J$ decreases (below 4\,K) with pressure.

The present work builds on the recent study by Ducatman {\it et al}~\cite{Sam2018}, which is based on the intuitive idea that the observed IC order~\cite{Biffin2014a} can be thought of as a long-wavelength twisting of a nearby commensurate order, called the `$K$-state' [see Fig.~\ref{fig:PT}\,(a)], with the same qualitative features. Namely, the same propagation vector (with periodicity very close to the experimental value $0.57$), the same irreducible representation, the counter-rotating moments, and non-coplanarity. The present study essentially addresses the fate of this $K$-state under a field along the ${\bf b}$-axis. But first, let us repeat the most important aspects of the system and the zero-field $K$-state.~\cite{Sam2018}

\vspace*{-0.5cm} 
\section{Lattice structure and microscopic spin model}
\begin{figure}[!t]
\includegraphics[width=1\columnwidth]{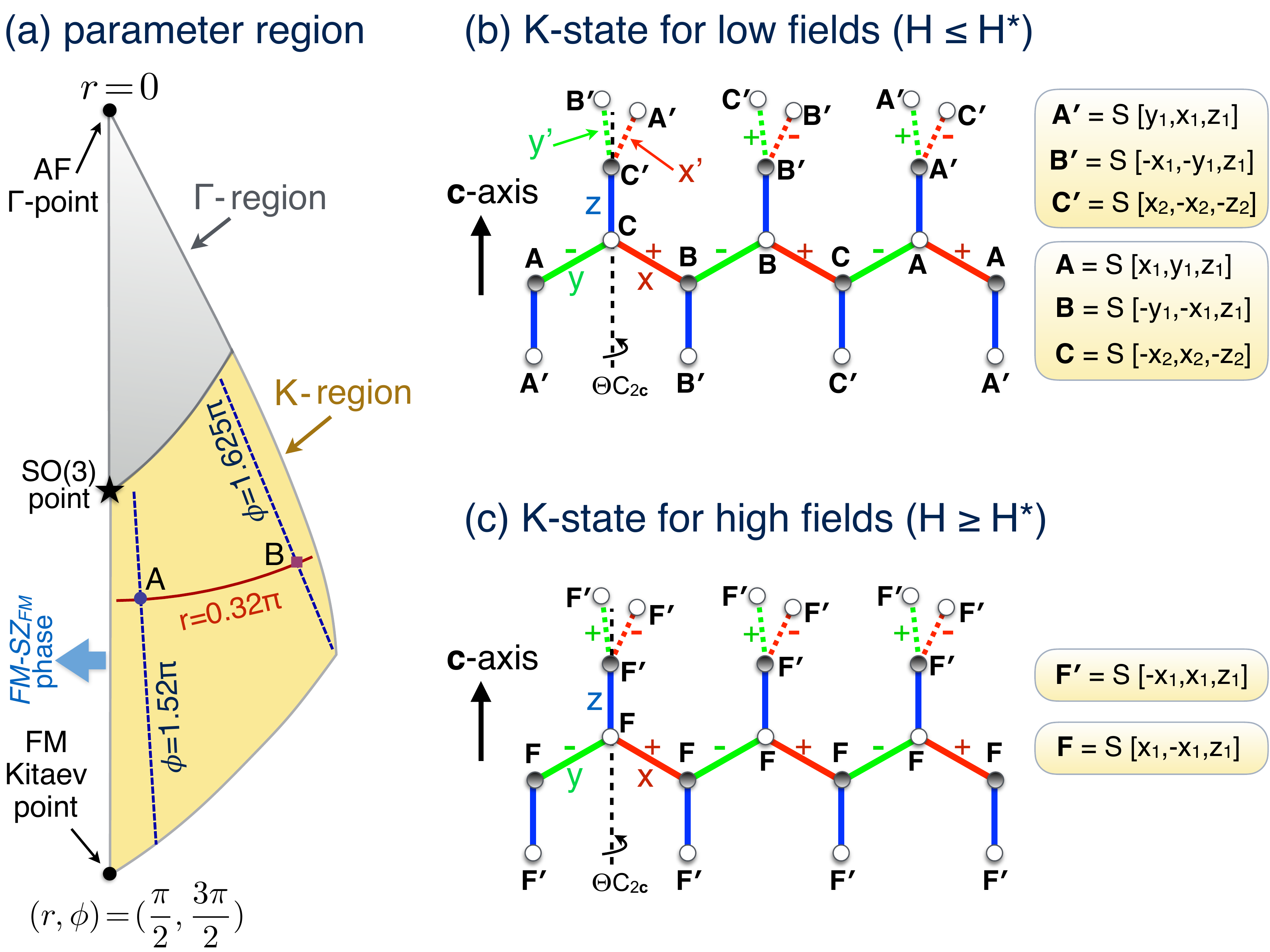}
\vspace*{-0.25cm}
\caption{
(a) The relevant parameter space in $(r,\phi)$ [see Eq.~(\ref{eq:rphi})], along with the $K$- and $\Gamma$-regions proposed in Ref.~[\onlinecite{Sam2018}]. 
The points $A$ and $B$, referred to in subsequent figures, correspond to $(r,\phi)\!=\!(0.32\pi,1.52\pi)$ and $(0.32\pi,1.625\pi)$. 
(b-c) The structure of the $K$-state at low ($H\!\le\!H^\ast$) and high ($H\!\ge\!H^\ast$) fields. The solid (dashed) green and red bonds depict the $xy$ ($x'y'$) chains running along $\hat{{\bf a}}$+$\hat{{\bf b}}$ ($\hat{{\bf a}}$-$\hat{{\bf b}}$). The blue vertical segments point along the ${\bf c}$-axis and depict the $z$-bonds.
The Cartesian components of the spins are shown in the side panels. 
Both states respect the combined operation $\Theta C_{2{\bf c}}$, where $\Theta$ is time reversal and $C_{2{\bf c}}$ is the two-fold rotation around ${\bf c}$. 
The high-field K-state is qualitatively similar to the state `FM-SZ$_{\text{FM}}$' of Ref.~[\onlinecite{Lee2015}], which is stabilized in a finite region with $\phi\!<\!3\pi/2$.
}
\label{fig:PT}
\end{figure}

\vspace*{-0.3cm}
The lattice structure of $\beta$-Li$_2$IrO$_3$ has been discussed in detail elsewhere.~\cite{Biffin2014a,Lee2015,Lee2016,Ruiz2017,Sam2018}
The Ir ions form interwoven networks of two types of `zig-zag' chains, one propagating along $\hat{{\bf a}}$+$\hat{{\bf b}}$ and the other along $\hat{{\bf a}}$-$\hat{{\bf b}}$. 
The first type of chains comprise the nearest-neighbor (NN) bonds denoted by $x$ and $y$ in Fig.~\ref{fig:PT}\,(b), while the second type comprise the NN bonds denoted by $x'$ and $y'$. We shall refer to these as $xy$- and $x'y'$-chains, respectively.
Finally, the two chain types are connected via NN bonds that are oriented along the $\hat{{\bf c}}$-axis and are denoted by $z$ in Fig.~\ref{fig:PT}\,(b).

The microscopic $J$-$K$-$\Gamma$ model, introduced by Lee {\it et al},~\cite{Lee2015,Lee2016} features three types of NN interactions, the Heisenberg exchange $J$, the Kitaev anisotropy $K$ and the off-diagonal symmetric anisotropy $\Gamma$. For a given bond of type $t$, between NN sites $i$ and $j$, the total interaction takes the form
\be
\mc{H}_{ij}^{(t)} = J \vec{S}_i\cdot \vec{S}_j
+ K S_i^{\alpha_t} S_j^{\alpha_t} 
+\sigma_t \Gamma (S_i^{\beta_t} S_j^{\gamma_t}+S_i^{\beta_t} S_j^{\gamma_t})\,,
\ee
where $(\alpha_t,\beta_t,\gamma_t)\!=\!(x,y,z)$ for $t\!\in\!\{x,x'\}$, 
$(y,z,x)$ for $t\!\in\!\{y,y'\}$, and $(z,x,y)$ for $t\!=\!z$. 
The prefactor $\sigma_t$ of the $\Gamma$ terms is $+1$ for $t\!\in\!\{x, z, y'\}$ and $-1$ for $t\!\in\!\{y,x'\}$.
This modulation of the prefactors is tied to the following convention for the relation between the crystallographic axes $\{\hat{{\bf a}}, \hat{{\bf b}},\hat{{\bf c}}\}$ and the Cartesian axes $\{\hat{{\bf x}}, \hat{{\bf y}}, \hat{{\bf z}}\}$~\footnote{From these relations it follows, in particular, that the $\hat{\bf b}$-axis is special because the Kitaev exchange on the $z$-type of bonds couples the spin projections along precisely this axis.~\cite{Ruiz2017}}:
\be
\label{eq:abc}
\begin{array}{c}
\hat{{\bf x}}=\frac{\hat{{\bf a}}+\hat{{\bf c}}}{\sqrt{2}}, ~
\hat{{\bf y}}=\frac{\hat{{\bf c}}-\hat{{\bf a}}}{\sqrt{2}}, ~
\hat{{\bf z}}=-\hat{{\bf b}}.
\end{array}
\ee
The total Hamiltonian in a field ${\bf H}$ takes the form
\be
\begin{array}{c}
\mc{H}=\sum_t \sum_{\langle ij\rangle \in t} \mc{H}_{ij}^{(t)} - \mu_B {\bf H} \cdot \sum_{i} {\bf g}_i \cdot {\bf S}_i\,,
\end{array}
\ee 
where $\mu_B$ is the Bohr magneton and ${\bf g}_i$ is the electronic ${\bf g}$-tensor at site $i$.
In the following, we work in units of $\sqrt{J^2\!+\!K^2\!+\!\Gamma^2}\!=\!1$ and use the parametrization~\cite{Lee2015,Lee2016,Sam2018}:
\be\label{eq:rphi}
\begin{array}{c}
J=\sin r \cos\phi, ~
K=\sin r \sin\phi, ~
\Gamma = -\cos r\,,
\end{array}\ee
where $\phi\!\in\![0,2\pi)$, $r\!\in\![0,\pi/2]$. 
In particular, we shall focus on the `$K$-region' of Fig.~\ref{fig:PT}\,(a). As argued in Ref.~\cite{Sam2018}, the actual IC order inside this region can be thought of as a long-wavelength twisting of a nearby commensurate state, called the `$K$-state'. This state is a local minimum of the energy, except at $\phi\!=\!3\pi/2$ where it is one of the global minima (which form an $\mc{S}^2$ manifold), and probably survives as such in a small finite window of $\phi$ above $3\pi/2$ due to the lattice cut-off.~\cite{Sam2018} Nevertheless, this state has all qualitative features of the experimentally observed IC order at zero field~\cite{Sam2018}.

\vspace*{-0.25cm} 
\section{Main aspects of the zero-field $K$-state}
\vspace*{-0.3cm}
The detailed structure of the $K$-state is discussed in Ref.~\cite{Sam2018}.  A schematic representation is shown in Fig.~\ref{fig:PT}\,(b). 
Let us repeat here the main features that are needed for our purposes.
The $K$-state features six spin sublattices, three (${\bf A}$, ${\bf B}$, ${\bf C}$) along the $xy$-chains and three (${\bf A}'$, ${\bf B}'$, ${\bf C}'$) along the $x'y'$-chains. The corresponding Cartesian components are given in the side panels of Fig.~\ref{fig:PT}\,(b), where $S=1/2$ is the classical spin length, while the numbers $x_1$, $y_1$, $z_1$, $x_2$ and $z_2$ are all positive (at zero field), and obey the constraints $x_1^2+y_1^2+z_1^2\!=\!1$ and $2x_2^2+z_2^2\!=\!1$. 

The sublattices $\{{\bf A}, {\bf B}, {\bf C}\}$ (and likewise the sublattices  $\{{\bf A}', {\bf B}', {\bf C}'\}$) form an almost ideal 120$^\circ$-pattern. 
The counter-rotating modulation of the moments can be seen in Fig.~\ref{fig:PT}\,(b) by noticing that, along the $xy$-chains (similarly for the $x'y'$-chains), the odd sites (gray circles) modulate in a ${\bf A B C A B} \cdots$ pattern, while the even sites (white circles) show a ${\bf C B A C B}\cdots$ pattern.
This modulation shows up in the Fourier component of the static structure factor at ${\bf Q}\!=\!2\hat{\bf a}/3$, which takes the characteristic form ${\bf M}_{{\bf Q}={2\hat{{\bf a}}/3}}\!=\!(i M_a A, i M_b C, M_c F)$, along ${\bf a}$, ${\bf b}$ and ${\bf c}$. Here, $A$, $C$ and $F$ denote, respectively, the N\'eel, stripy and FM basis vectors of the 4-site unit cell~\cite{Biffin2014a,Sam2018}. The amplitudes $M_a$, $M_b$ and $M_c$ are given by~\cite{Sam2018}
\be\label{eq:SofQ2o3}
\begin{array}{c}
M_a=i 2S (x_1+2x_2-y_1),~~M_b=-i2S (z_1+z_2),\\
M_c=i 2S \sqrt{3}(x_1+y_1)\,.
\end{array}
\ee
The counter-rotating modulation is however not the only component of the $K$-state, because there are two types of deviations from the ideal 120$^\circ$-pattern when $\phi\!>\!3\pi/2$: 
i) an in-plane canting of the zigzag type, whose direction alternates between ${\bf a}$ and $-{\bf a}$ for $xy$- and $x'y'$-chains, respectively, and ii) an out-of-plane ferromagnetic (FM) canting along ${\bf b}$. 
Both of these cantings are uniform from one unit cell to another, and show up directly in the ${\bf Q}=0$ Fourier component of the static spin structure factor, along ${\bf a}$ and ${\bf b}$. Specifically, ${\bf M}_{{\bf Q}=0}\!=\!(M_a' G, M_b' F, 0)$, where $G$ and $F$ denote, respectively, the zigzag and FM basis vectors of the 4-site primitive unit cell~\cite{Biffin2014a,Sam2018}. 
The amplitudes $M_a'$ and $M_b'$ are~\footnote{Here and in Eq.~(\ref{eq:SofQ2o3}) we have multiplied by a factor of $2S$ compared to the expressions given in Ref.~[\onlinecite{Sam2018}].}  
\be\label{eq:SofQ0}
M_a' = -4S(x_1-y_1-x_2), ~~ M_b' = 2S (2z_1-z_2)\,.
\ee
These amplitudes vanish when $J\to 0^+$, because in this limit $\{{\bf A}, {\bf B}, {\bf C}\}$ and $\{{\bf A}', {\bf B}', {\bf C}'\}$ reach their ideal 120$^\circ$-patterns.

\vspace*{-0.25cm} 
\section{Total Energy of the $K$-state in a field}
\vspace*{-0.3cm}
We now move to the main part of this study and analyze the fate of the $K$-state under a field $H$ along the ${\bf b}$-axis. 
As discussed in detail in Ref.~\cite{Ruiz2017}, the total polarization along ${\bf b}$ couples to the field via the diagonal element $g_{bb}$ of the ${\bf g}$-tensor, while the uniform zigzag canting along ${\bf a}$ (i.e., the staggered magnetization from $xy$- to $x'y'$-chains), couples to the field via the off-diagonal element $g_{ab}$, whose sign alternates between $xy$ and $x'y'$ chains due to the two-fold symmetry $C_{2{\bf a}}$ that passes through the middle of the $z$ bonds.  
So, a magnetic field along ${\bf b}$ couples linearly to both $M_a'$ and $M_b'$.  
Furthermore, such a field does not break the symmetry $\Theta C_{2{\bf c}}$ obeyed by the $K$-state, see Fig.~\ref{fig:PT}\,(b). 
These arguments suggest that we can use the $K$-state ansatz [side panel of Fig.~\ref{fig:PT}\,(b)], but now the coefficients $x_1$, $y_1$, etc will change with the field. 
To find these coefficients we must minimize the total energy,
\bea\label{eq:En}
\begin{array}{l}
E/N \!=\! S^2\Big\{
K \left[ 3-2(y_1-x_2)^2 \right] \\
~~~\!+\!2 \Gamma \left[1-z_1^2 + x_2^2  + 2 (y_1+x_2) z_1 + 2 x_1 z_2 \right] \\ 
~~~\!+\! J \left[1+2 (z_1-z_2)^2- 4 x_1 x_2  + 4 (x_1+x_2) y_1 \right] \Big\}/6 \\
~~~
\!-\! \mu_B H S \left[\sqrt{2} g_{ab} (x_1 - x_2 - y_1) + g_{bb} (-2 z_1 + z_2)\right]/3\,,
\end{array}
\eea
($N$ is the total number of sites) for given $H$, $J$, $K$, $\Gamma$, $g_{bb}$, and $g_{ab}$. From $x_1$, $y_1$, etc we can then deduce the various components of the structure factor using Eqs.~(\ref{eq:SofQ2o3}) and (\ref{eq:SofQ0}).


\begin{figure}[!t]
\includegraphics[width=0.9\columnwidth]{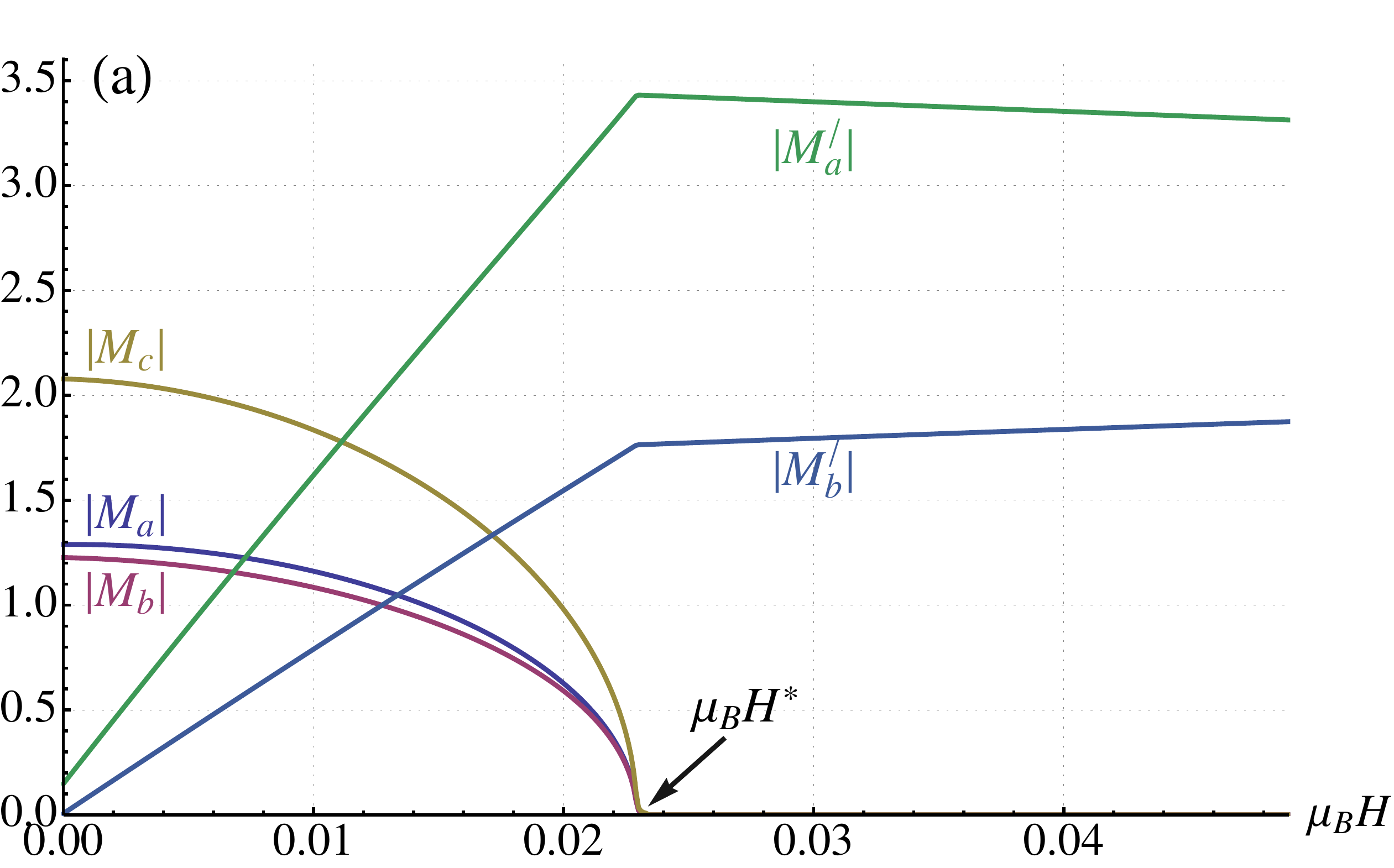}
\includegraphics[width=0.9\columnwidth]{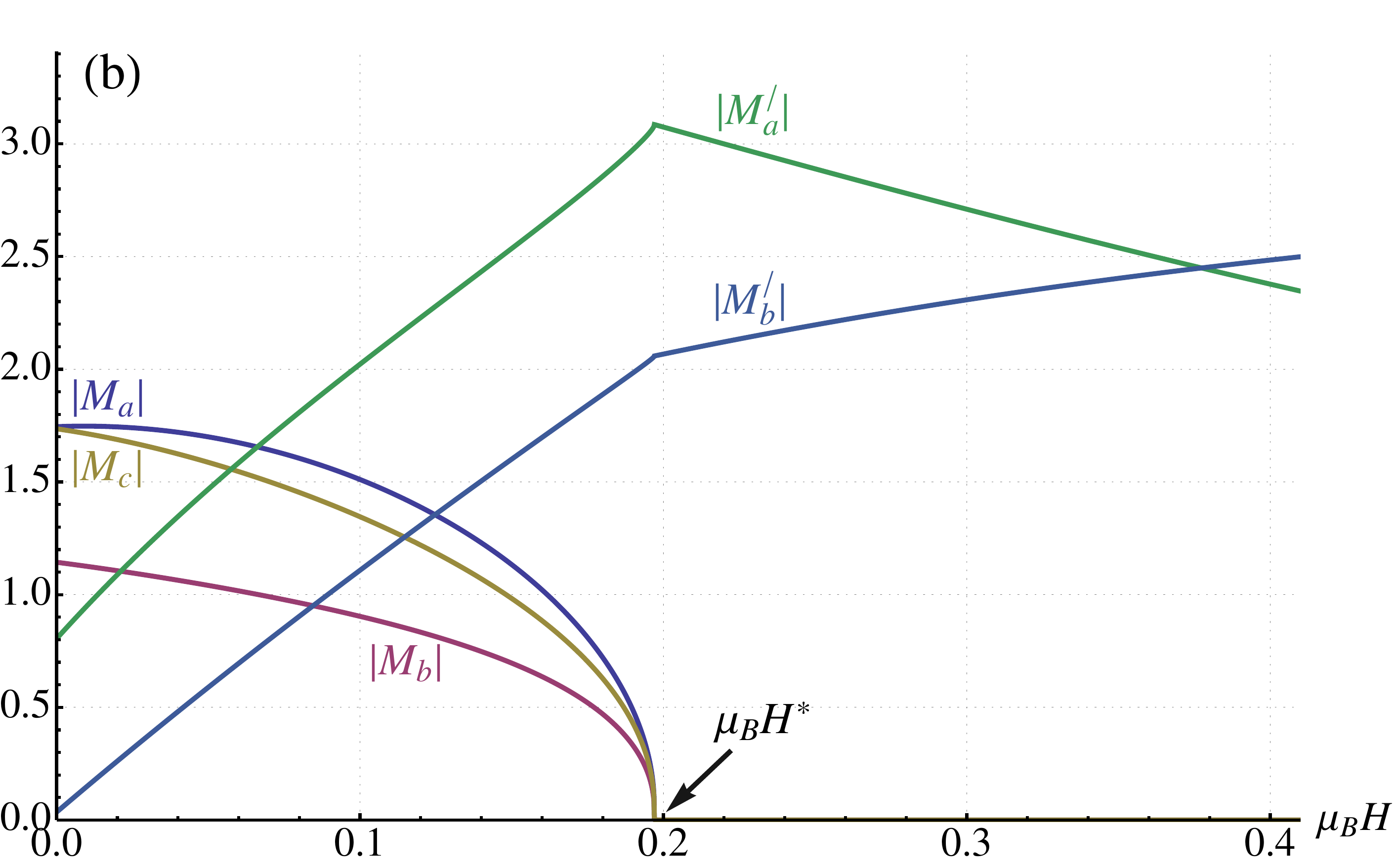}
\vspace*{-0.25cm}
\caption{Evolution of the various Fourier components of the static structure factor of the $K$-state with a magnetic field $H$ along ${\bf b}$, for the points $A$ (panel a) and $B$ (panel b) in Fig.~\ref{fig:PT}\,(a).}
\vspace*{-0.5cm}
\label{fig:SofQvsH}
\end{figure}

\vspace*{-0.25cm} 
\section{Main results}
\vspace*{-0.3cm}
Let us first highlight the results that are directly related to the reported experiments.
For demonstration, we have taken $g_{bb}\!=\!2$ and $g_{ab}\!=\!0.1$ (we shall address the role of $g_{ab}$ separately below).
Figs.~\ref{fig:SofQvsH}\,(a) and (b) show the magnitudes of the various Fourier components, at the points $A$ and $B$ of Fig.~\ref{fig:PT}\,(a). 
The results show that the ${\bf Q}\!=\!2\hat{{\bf a}}/3$ components, $M_a$, $M_b$ and $M_c$, decline with the field, and vanish completely at (and above) a characteristic field $H^\ast$. 
At the same time, the ${\bf Q}\!=\!0$ components, $M_a'$ and $M_b'$, grow linearly with the field, and they show a kink at $H^\ast$. These results are consistent with the data reported in Ref.~[\onlinecite{Ruiz2017}].

We can also see an important feature of the ${\bf Q}\!=\!0$ components, which is related to the zero-field scattering experiments of Biffin {\it et al}~\cite{Biffin2014a}.
Namely, that both $|M_a'(0)|$ and $|M_b'(0)|$ are very small when $\phi$ is close to $3\pi/2$ [see Fig.~\ref{fig:SofQvsH}\,(a)], where both the zigzag and the FM canting become small. 
As we move away from the line $\phi\!=\!3\pi/2$, the zigzag amplitude $M_a'(0)$, in particular, is not small any longer, see Fig.~\ref{fig:SofQvsH}\,(b). 
So, the absence of the ${\bf Q}\!=\!0$ Bragg peaks from the zero-field scattering experiments of Biffin {\it et al}.~\cite{Biffin2014a} is the first evidence that $\phi$ is close to $3\pi/2$, i.e., that $J$ is much weaker than both $K$ and $\Gamma$.

The next qualitative experimental result that we would like to address is the intensity sum rule of Ref.~[\onlinecite{Ruiz2017}], i.e. the finding that the sum of a certain combination of the intensities corresponding to zero- and finite-$Q$ Bragg peaks, remains almost constant even slightly above $H^\ast$.
To this end, we will need to understand the behavior of the coefficients $x_1$, $y_1$, etc first. 

\begin{figure}[!t]
\includegraphics[width=0.95\columnwidth]{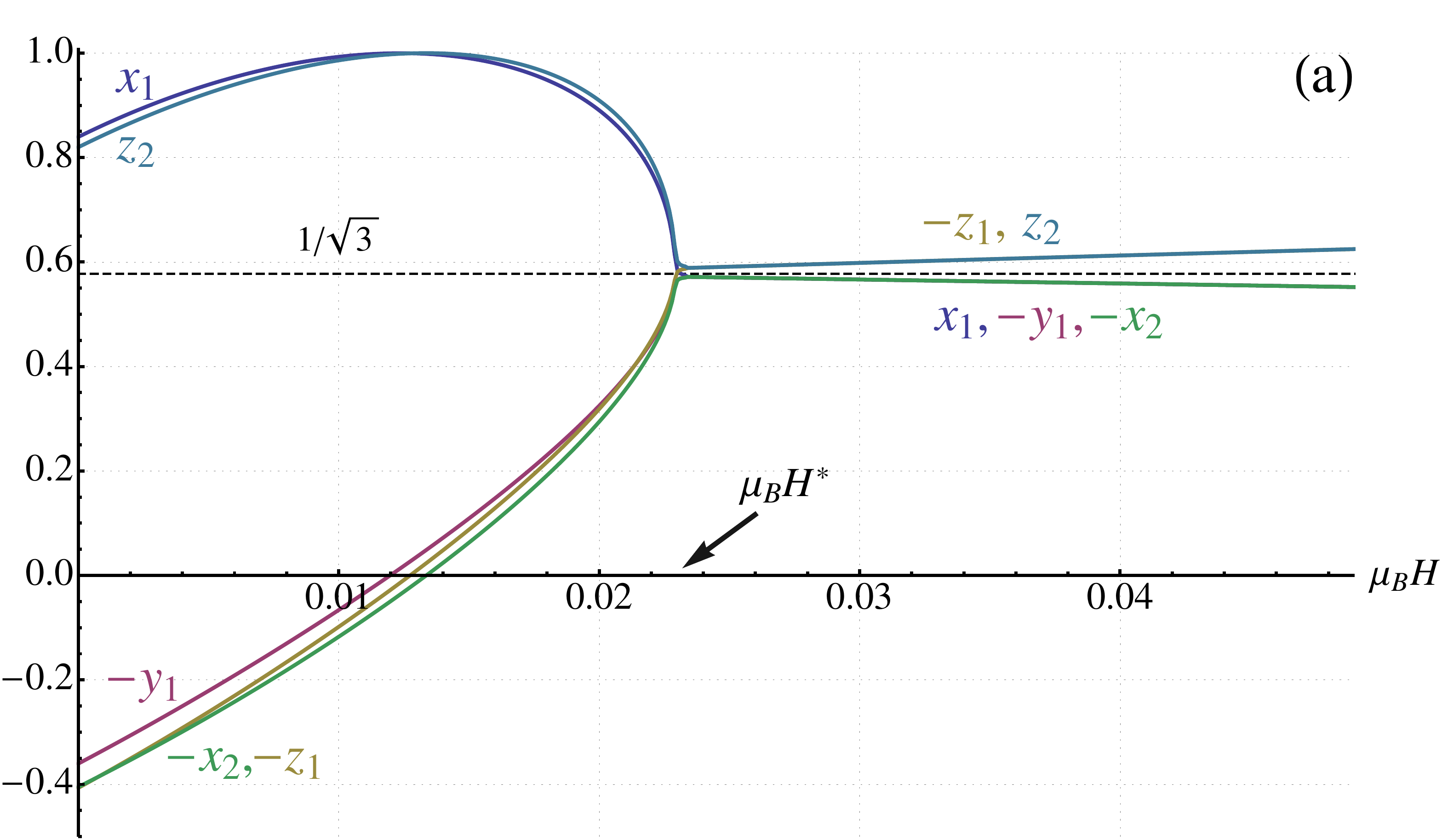}
\includegraphics[width=0.95\columnwidth]{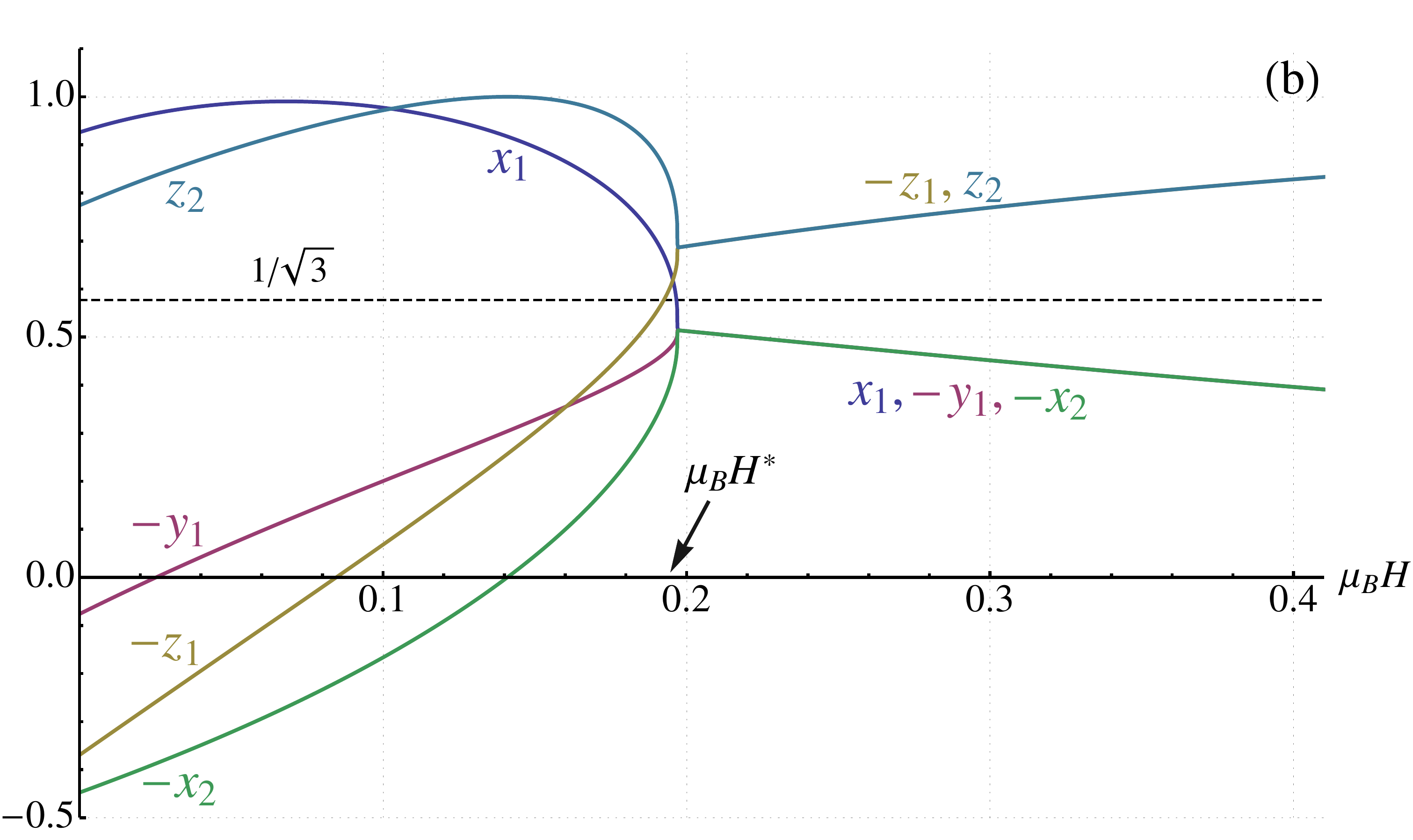}
\caption{Evolution of the coefficients $x_1$, $y_1$, etc with a magnetif field $H$ along ${\bf b}$, for the points $A$ (panel a) and $B$ (panel b) in Fig.~\ref{fig:PT}\,(a)}
\vspace*{-0.25cm}
\label{fig:x1y1etcvsH}
\end{figure}

From Eqs.~(\ref{eq:SofQ2o3}-\ref{eq:SofQ0}) we find that the simultaneous vanishing of $M_a$, $M_b$ and $M_c$ for $H\!\ge\!H^\ast$ imply that 
\be\label{eq:HgeHc}
H\ge H^\ast:~~x_1=-y_1=-x_2, ~~
z_2=-z_1\,.
\ee
These relations are indeed satisfied as we see in Fig.~\ref{fig:x1y1etcvsH}. In particular, while all coefficients are positive at zero field, the coefficients $y_1$, $x_2$ and $z_1$ change sign at some intermediate field, and eventually satisfy (\ref{eq:HgeHc}) for $H\ge H^\ast$.

The explanation of the intensity sum rule reported in Ref.~[\onlinecite{Ruiz2017}] stems from another aspect of Fig.~\ref{fig:x1y1etcvsH}\,(a), namely, that 
\be\label{eq:Approx}
\begin{array}{l}
H\!\le\!H^\ast,\\
 \phi\to\left(\frac{3\pi}{2}\right)^+
\end{array}\!\!:
\left\{\!\!
\begin{array}{l}
z_2\simeq x_1,~~x_2\simeq z_1\simeq y_1,\\
x_1(0)\simeq\sqrt{2/3},~~y_1(0)\simeq1/\sqrt{6}.
\end{array}\right.
\ee
Remarkably, these approximate relations hold all the way down to zero field, where they stem from the special structure of the ground state manifold along $\phi\!=\!3\pi/2$, and the concomitant lifting of the degeneracy  by an infinitesimal positive $J$.~\cite{Sam2018}  
For larger $J$, the approximations in Eq.~(\ref{eq:Approx}) become progressively worse as shown in Fig.~\ref{fig:x1y1etcvsH}\,(b).

\begin{figure}[!t]
\includegraphics[width=0.9\columnwidth]{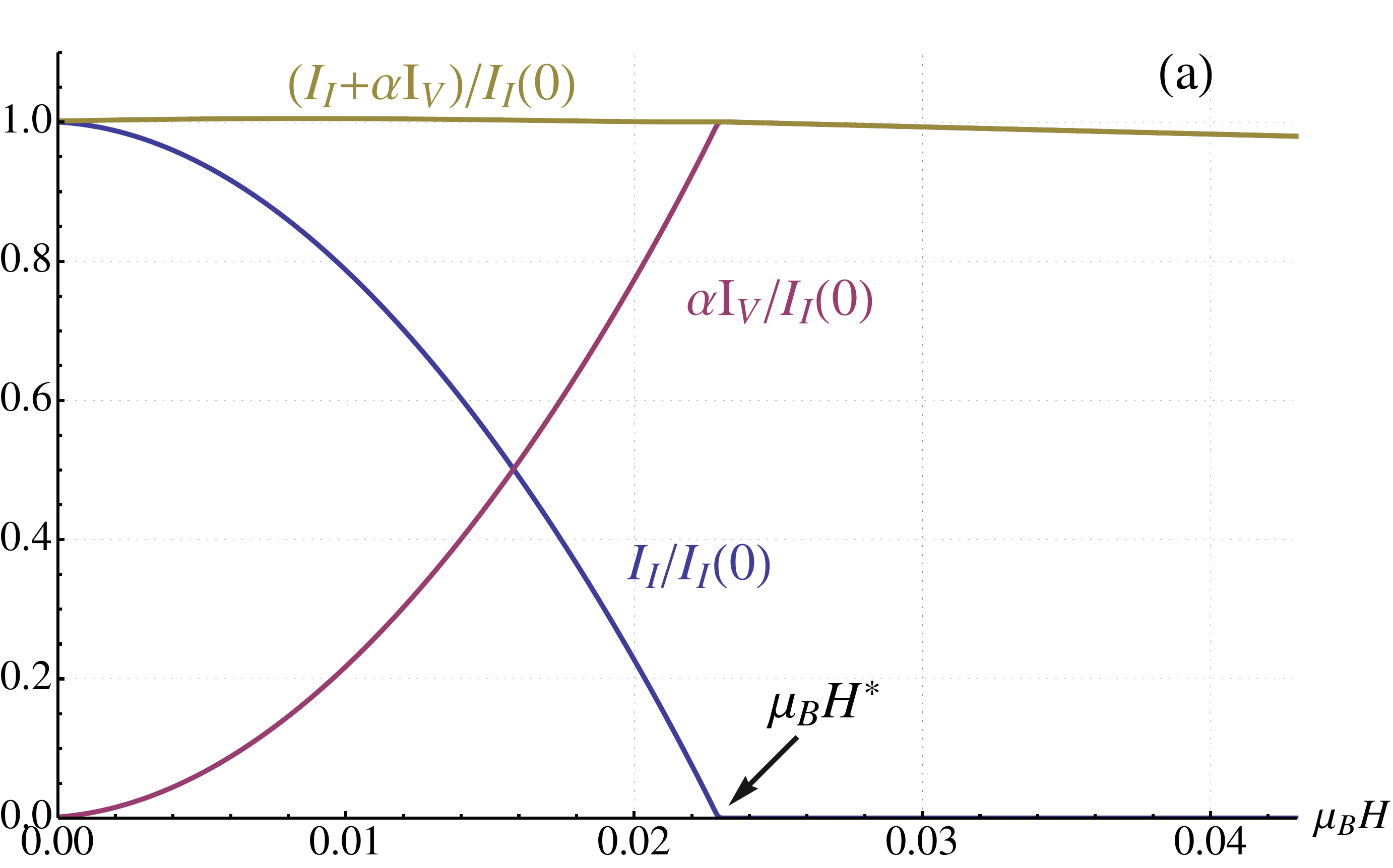}
\includegraphics[width=0.9\columnwidth]{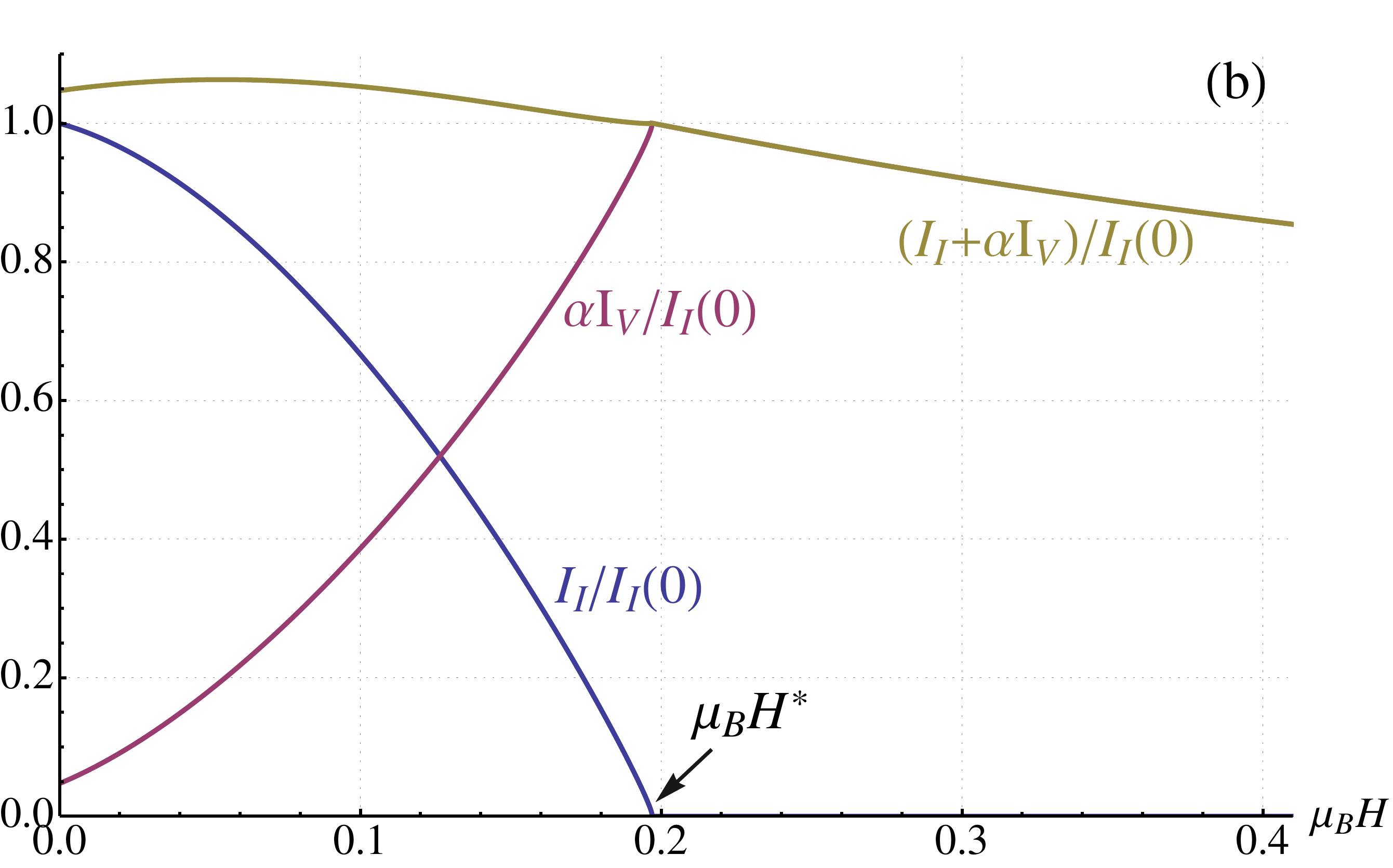}
\caption{Evolution of the combinations $I_I$, $I_V$ and $I_{\text{tot}}$ [see Eq.~(\ref{eq:SumRule})] with a magnetic field $H$ along ${\bf b}$, for the points $A$ (panel a) and $B$ (panel b) of Fig.~\ref{fig:PT}\,(a). 
In (a), $\alpha\!\simeq\!0.503$ [very close to the expected value of $1/2$ when $\phi\!\to\!(3\pi/2)^+$, see text], while in (b), $\alpha\!\simeq\!0.535$.}
\vspace*{-0.25cm}
\label{fig:SumRule}
\end{figure}

Now, to see how Eq.~(\ref{eq:Approx}) leads to the intensity sum rule, we take the following combinations of the Bragg peak intensities,  
\be\label{eq:SumRule}
\begin{array}{l}
I_{I}\!=\!|M_a|^2\!+\!|M_b|^2\!+\!|M_c|^2, ~~
I_{V}\!=\!|M_a'|^2\!+\!|M_b'|^2\,.
\end{array}
\ee
Fig.~\ref{fig:SumRule} shows the behavior of the quantities $I_I/I_I(0)$, $\alpha I_V/I_I(0)$ and $I_{\text{tot}}\!\equiv\!(I_I+\alpha I_V)/I_I(0)$, where the constant $\alpha\!\equiv\!I_I(0)/I_V(H^\ast)$ fixes $I_{\text{tot}}(H^\ast)\!=\!1$. 
We first discuss the results shown in Fig.~\ref{fig:SumRule}\,(a), which are obtained at the point $A$ of Fig.~\ref{fig:PT}\,(a). 
As expected, the intensity associated with the counter-rotating component of the order, $I_I$, declines quickly with field, while the intensity associated with the uniform components, $I_V$, grows quadratically with field up to $H^\ast$. 
At the same time, the total intensity $I_{\text{tot}}$ remains extremely close to $1$ from zero field all the way up to $H^\ast$. This behavior is fully consistent with the intensity sum rule of Fig.~4\,(a) of Ref.~[\onlinecite{Ruiz2017}].

Importantly, while it is not clear which exact combinations of the zero- and finite-${\bf Q}$ intensities are involved in the sum rule reported in Ref.~[\onlinecite{Ruiz2017}], we can show that this does not matter as long as we are close to the line $\phi\!=\!3\pi/2$, where the approximate relations (\ref{eq:Approx}) hold. Indeed, these relations give:
\be\label{eq:HleHc}
\begin{array}{l}
H\!\le\!H^\ast,\\
 \phi\to\left(\frac{3\pi}{2}\right)^+
 \end{array}\!\!:
\!\left\{\!\!
\begin{array}{l}
|M_a|^2 \!\simeq\! |M_b|^2 \!\simeq\! \frac{1}{3}|M_c|^2 \!\simeq\! (x_1\!+\!y_1)^2,\\
|M_b'|^2 \!\simeq\! \frac{1}{4}|M_a'|^2 \!\simeq\! (x_1\!-\!2y_1)^2.
\end{array}
\right.
\ee
This implies that any linear combination among $\{|M_a|^2$, $|M_b|^2$, $|M_c|^2$, $|M_a'|^2$, $|M_b'|^2\}$ will always be of the form $ (x_1\!+\!y_1)^2$+$\beta(x_1\!-\!2y_1)^2$, which, in turn, becomes independent of $H$ if we choose $\beta\!=\!1/2$ (using the spin length constraint $1\!=\!x_1^2\!+\!y_1^2\!+\!z_1^2\!\simeq \!x_1^2\!+\!2y_1^2$). This value is consistent with the limiting value of $\alpha$ defined above when $\phi\!\to\!(3\pi/2)^+$.

The intensity sum rule is no longer satisfied as we go further away from $\phi\!=\!3\pi/2$. This can be seen in Fig.~\ref{fig:SumRule}\,(b), which shows $I_I$, $I_V$ and $I_{\text{tot}}$ computed at the point $B$ of Fig.~\ref{fig:PT}\,(a). The total intensity $I_{\text{tot}}$ deviates from the value $1$ in the entire region between $H\!=\!0$ and $H\!=\!H^\ast$, except at $H^\ast$ where it is fixed to $1$ by definition. 
We also see that $I_{\text{tot}}$ shows a fast drop below $1$ at $H\!>\!H^\ast$, in contrast to the experimental data,~\cite{Ruiz2017} but also in contrast to the data of Fig.~\ref{fig:SumRule}\,(a), where the corresponding drop at $H\!>\!H^\ast$ is much slower. 
Note also that the overall deviation from the sum rule could  be even worse for linear combinations of $\{|M_a|^2$, $|M_b|^2$, $|M_c|^2$, $|M_a'|^2$, $|M_b'|^2\}$ other that the ones involved in $I_I$ and $I_V$.  
We can therefore safely conclude that the experimental observation of the sum rule is an independent signature of the smallness of $J$.

\vspace*{-0.25cm}
\section{Nature of the high-field state ($H\ge H^\ast$)}
\vspace*{-0.3cm}
Based on Eq.~(\ref{eq:HgeHc}), the Cartesian components of the various spin sublattices for $H\!\ge\!H^\ast$ are:
\be\label{eq:StateHc}
H\ge H^\ast:
\left\{\!\!
\begin{array}{l}
{\bf A}={\bf B}={\bf C}=S [x_1,-x_1,z_1],\\
{\bf A}'={\bf B}'={\bf C}'= S [-x_1,x_1,z_1],
\end{array}
\right.
\ee
see Fig.~\ref{fig:PT}\,(c).  
So the $xy$- and $x'y'$-chains form two separate FM subsystems, which give a total FM moment along $-{\bf z}\!=\!{\bf b}$ [$z_1(H^\ast)$ is negative] and a staggered, zigzag moment along $\frac{{\bf x}-{\bf y}}{\sqrt{2}}\!=\!{\bf a}$ [see Eq.~(\ref{eq:abc})].  
This state is qualitatively the same with the state `FM-SZ$_{\text{FM}}$' reported by Lee {\it et al}~\cite{Lee2015,Lee2016} for $\phi\!<\!3\pi/2$, see Fig.~\ref{fig:PT}\,(a). 
The only difference is that in that state $x_1\!=\!\frac{1}{\sqrt{3}}$, while here $x_1$ in general deviates from this value and depends on the field. 
However, $x_1$ becomes very close to $\frac{1}{\sqrt{3}}$ precisely at $H\!=\!H^\ast$ when $\phi\!\to\!(3\pi/2)^+$, see Fig.~\ref{fig:x1y1etcvsH}\,(a).  
So, the effect of the field is to suppress the counter-rotating component of the state  and, at the same time, effectively drive the system towards the state that is stabilized for negative $J$. Intuitively then, the field plays the role of a FM Heisenberg coupling that counteracts the effect of $J$ (which is positive). 
This also tells us that the field $H^\ast$ required to achieve this must be proportional to $J$, and we will confirm this below. 

Now, the approximate relation $x_1(H^\ast)\!\approx\!\frac{1}{\sqrt{3}}$ for $\phi\!\to\!(2\pi/3)^+$ gives, based on Eq.~(\ref{eq:SofQ0}), $M_a'(H^\ast)\approx -2\sqrt{3}$ and $M_b'(H^\ast)\approx -\sqrt{3}$, see Fig.~\ref{fig:SofQvsH}\,(a). Remarkably, these values are independent of the ratio $K/\Gamma$, as long as we are inside the $K$-region of Fig.~\ref{fig:PT}\,(a) and close to the line $\phi\!=\!3\pi/2$. 
Furtermore, imposing (\ref{eq:HgeHc}) to (\ref{eq:En}) and minimizing the resulting expression for $E/N$ gives an implicit relation for $x_1(H)$:
\be\label{eq:Hvsx1}
\begin{array}{c}
H\ge H^\ast:~~\mu_B H = 2S \frac{\Gamma(4x_1^2-1)+(2J-\Gamma)x_1 \sqrt{1-2x_1^2}}{2g_{bb} x_1-\sqrt{2} g_{ab}\sqrt{1-2x_1^2}}\,.
\end{array}
\ee
Note that $K$ does not appear explicitly in this relation, which can give the wrong impression that $H^\ast$ does not depend on $K$. This is however not true, because 
$H^\ast$ and $x_1(H^\ast)$ 
cannot be both determined solely by Eq.~(\ref{eq:Hvsx1}).

Eq.~(\ref{eq:Hvsx1}) gives the large-field behavior of $M_a'$ and $M_b'$,
\be
H\ge H^\ast: ~~ M_a'=-12S x_1, ~~M_b' = - 6S \sqrt{1-2x_1^2}\,.
\ee
The large-field behavior of $x_1$, $|M_a'|$ and $|M_b'|$ are shown in Figs.~\ref{fig:x1y1etcvsH_WholeRegion} and \ref{fig:SofQvsH_WholeRegion}.
In the infinite-field limit, in particular, 
\be
\begin{array}{c}
H\!\to\!\infty:
~
M_a' \!\to\! -\frac{6g_{ab}}{\sqrt{2}\sqrt{g_{ab}^2+g_{bb}^2}},
~
M_b ' \!\to\! -\frac{3g_{bb}}{\sqrt{g_{ab}^2+g_{bb}^2}}\,.
\end{array}
\ee
With $g_{ab}\!\ll\!g_{bb}$ we get $M_a'\!\simeq\!0$,  $M_b'\!\simeq\!-3$, $x_1\!\simeq\!0$, $z_1\!\simeq\!-1$, and ${\bf A}\!=\!{\bf B}\!=\!{\bf C}\!=\!{\bf A}'\!=\!{\bf B}'\!=\!{\bf C}'\!=\!S \hat{{\bf b}}$, as expected.

\vspace*{-0.25cm}
\section{The role of $g_{ab}$ and the origin of the zigzag component}
\vspace*{-0.3cm}
As announced in the Introduction, the significant growth of the zigzag component under the field along ${\bf b}$ does not originate in the linear coupling between the zigzag component and the field, via $g_{ab}$. 
To show this we compare the response shown already in Fig.~\ref{fig:SofQvsH}\,(a) for $g_{ab}\!=\!0.1$, with the response for $g_{ab}\!=\!0$. 
This comparison shows that, while $H^\ast$ shifts to slightly higher value, the large size of the zigzag component at $H^\ast$ remains robust. 
A further comparison to the case of $g_{ab}\!=\!-0.1$ shows that even the choice of the sign of $g_{ab}$, which has been arbitrarily considered to be positive so far [i.e., positive (negative) on the $xy$ ($x'y'$) bonds], does not alter the large magnitude of $|M_a'(H^\ast)|$. 
Taken together, these results show that, although $M_a'$ couples linearly to the field via $g_{ab}$, its significant growth is not related to $g_{ab}$ but to an inherent tendency of the system to reach the state described by Eq.~(\ref{eq:StateHc}). This state has already a finite component at zero-field, although its amplitude is undetectable for weak $J$. 

\begin{figure}[!t]
\includegraphics[width=0.49\textwidth]{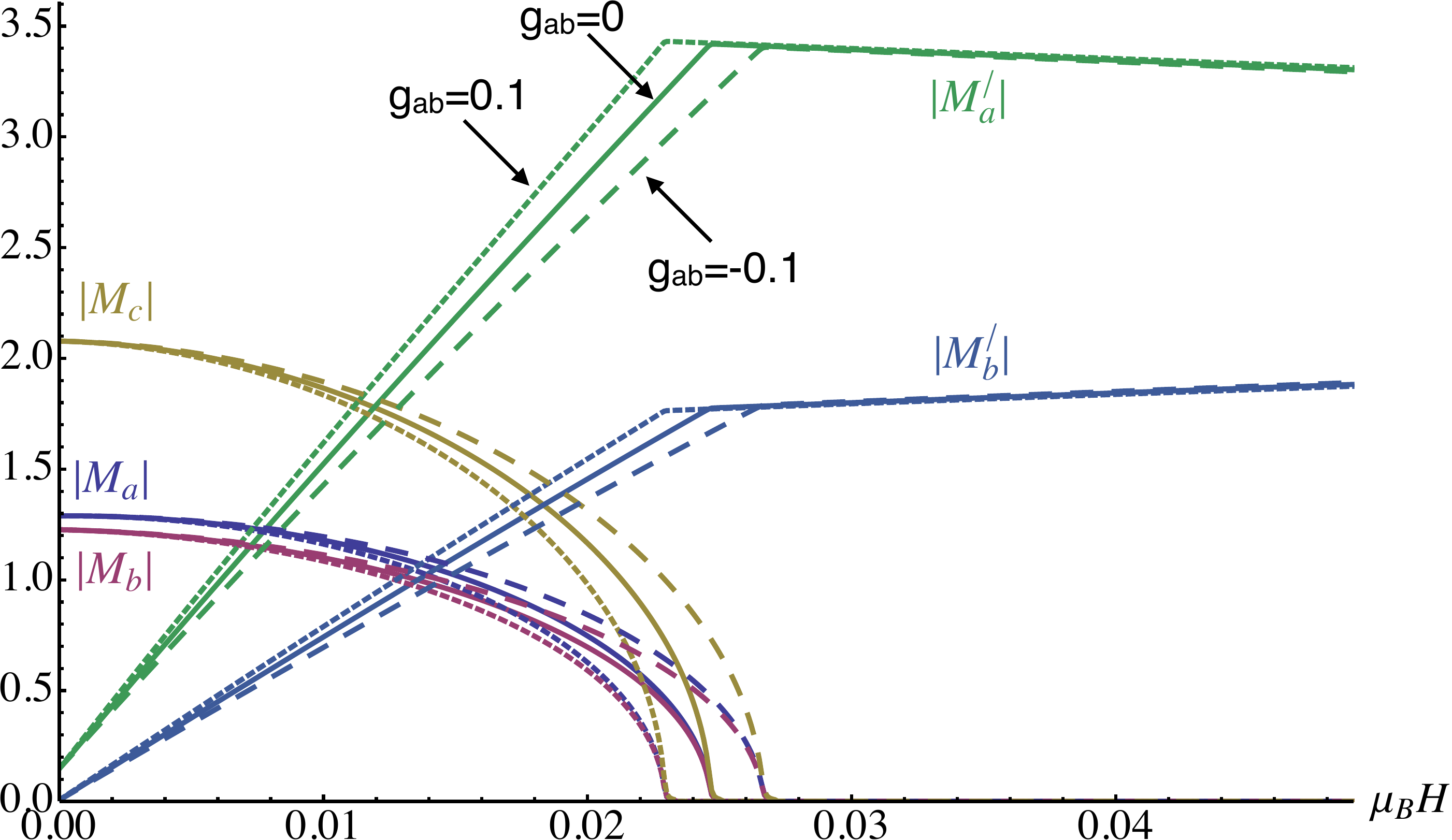}
\caption{
Effect of $g_{ab}$ on the evolution of the various components of the structure factor. 
Data are taken at the point $A$ of Fig.~\ref{fig:PT}\,(a), with   $g_{bb}\!=\!2$ and $g_{ab}\!=\!0.1$ [dotted lines, same as in Fig.~\ref{fig:SofQvsH}\,(a)], 0 (solid) and $-0.1$ (long-dashed).}
\vspace*{-0.25cm}
\label{fig:gab}
\end{figure}

\vspace*{-0.3cm}
\section{Dependence of $H^\ast$ on model parameters}
\begin{figure}[!t]
\includegraphics[width=0.49\textwidth]{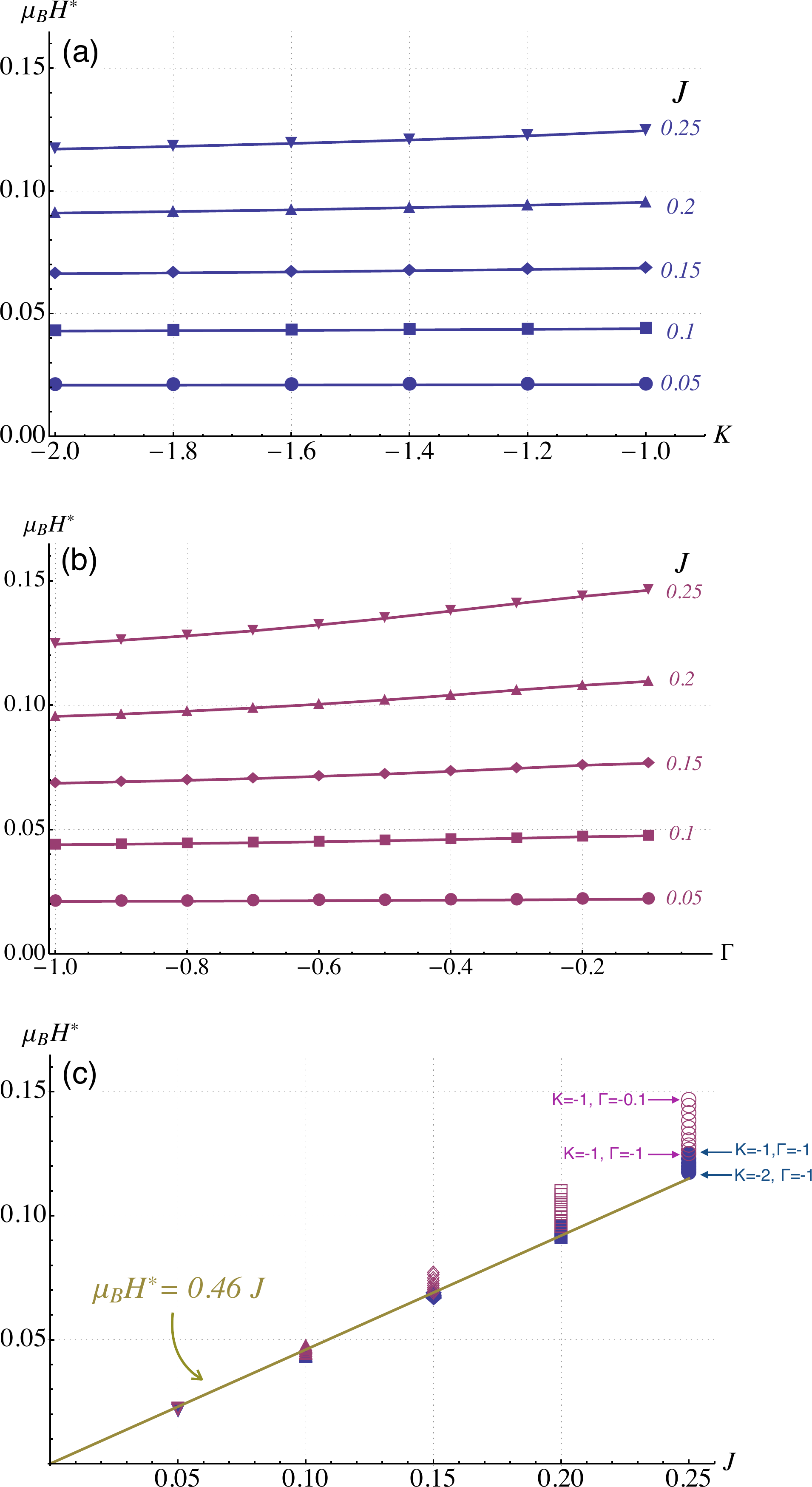}
\caption{
(a) Evolution of $H^\ast$ as a function of $K$ for fixed $\Gamma\!=\!-1$ and various values of $J$. 
(b) Evolution of $H^\ast$ as a function of $\Gamma$ for fixed $K\!=\!-1$ and various values of $J$. 
(c) Collected data shown in panels (a) and (b), with $J$ on the horizontal axis.  The line shown in a guide to the eye. 
For all cases we have taken $g_{ab}\!=\!0.1$ and $g_{bb}\!=\!2$.}
\vspace*{-0.25cm}
\label{fig:HcFixed}
\end{figure}

\begin{figure}[!t]
\includegraphics[width=0.9\columnwidth]{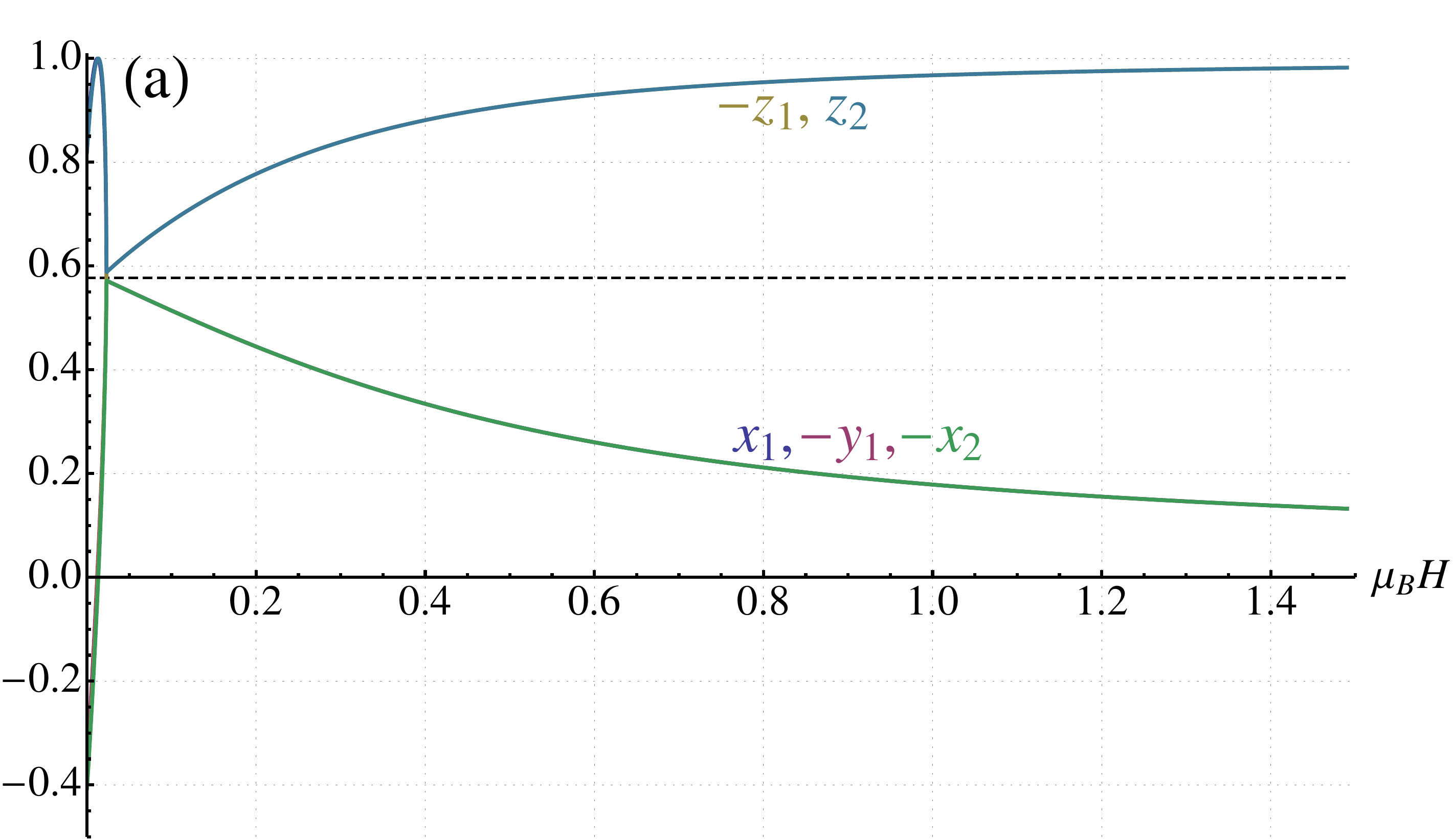}
\includegraphics[width=0.9\columnwidth]{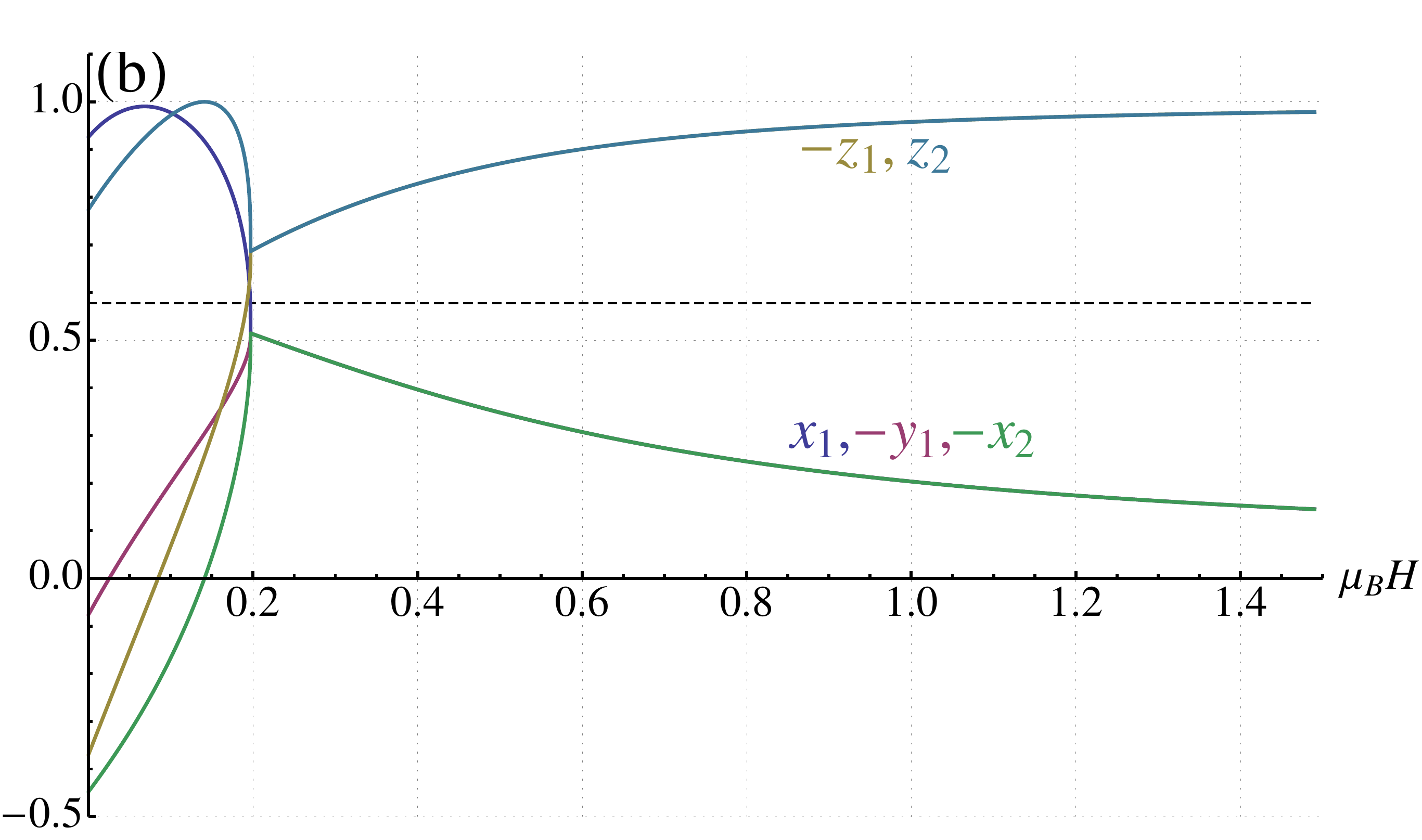}
\caption{Same as in Fig.~\ref{fig:x1y1etcvsH} but now we show the behavior up to fields much higher than $H^\ast$. Panels (a) and (b) correspond again to the parameter points $A$ and $B$, respectively, of Fig.~\ref{fig:PT}\,(a).}
\vspace*{-0.25cm}
\label{fig:x1y1etcvsH_WholeRegion}
\end{figure}

\begin{figure}[!t]
\includegraphics[width=0.9\columnwidth]{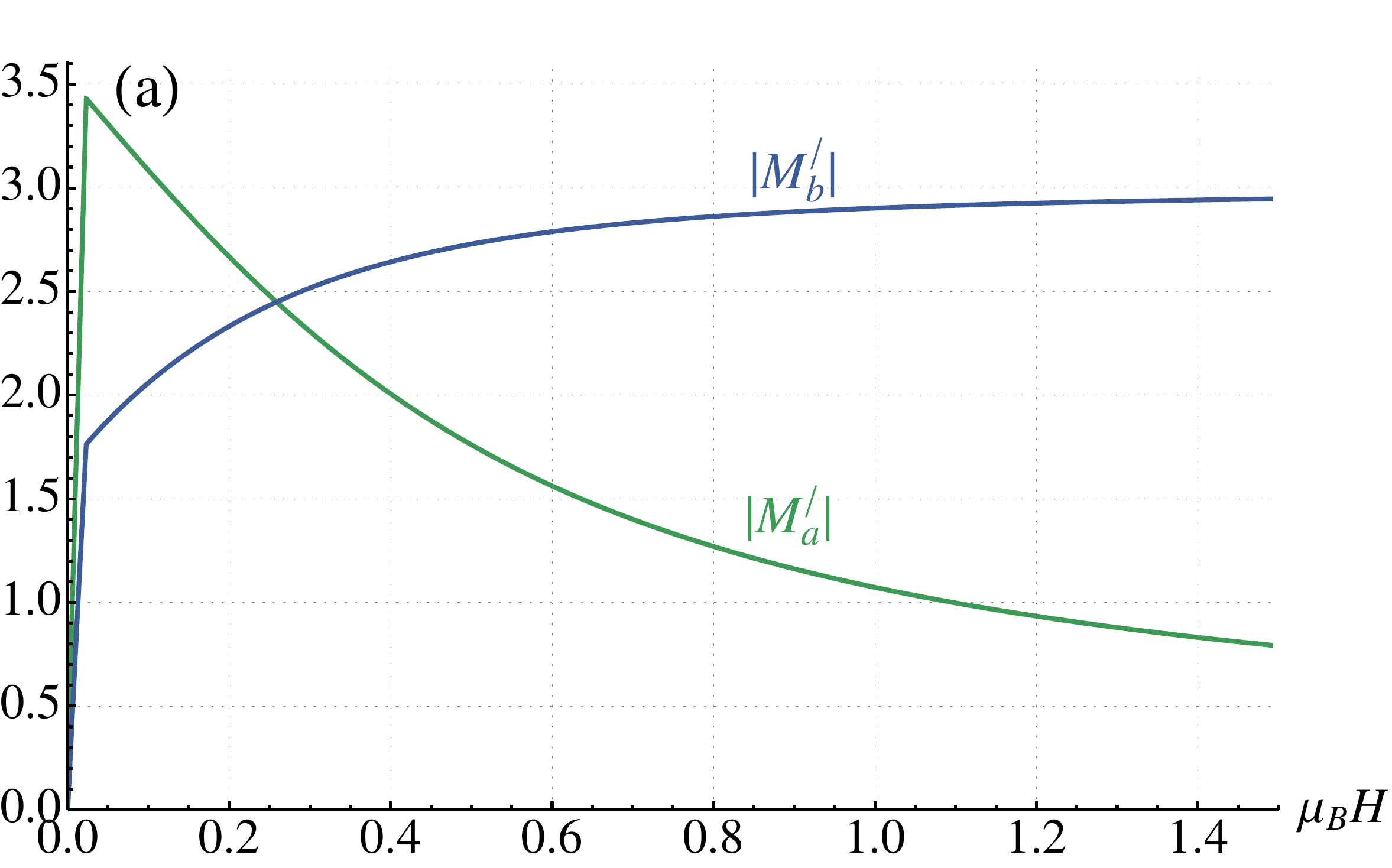}
\includegraphics[width=0.9\columnwidth]{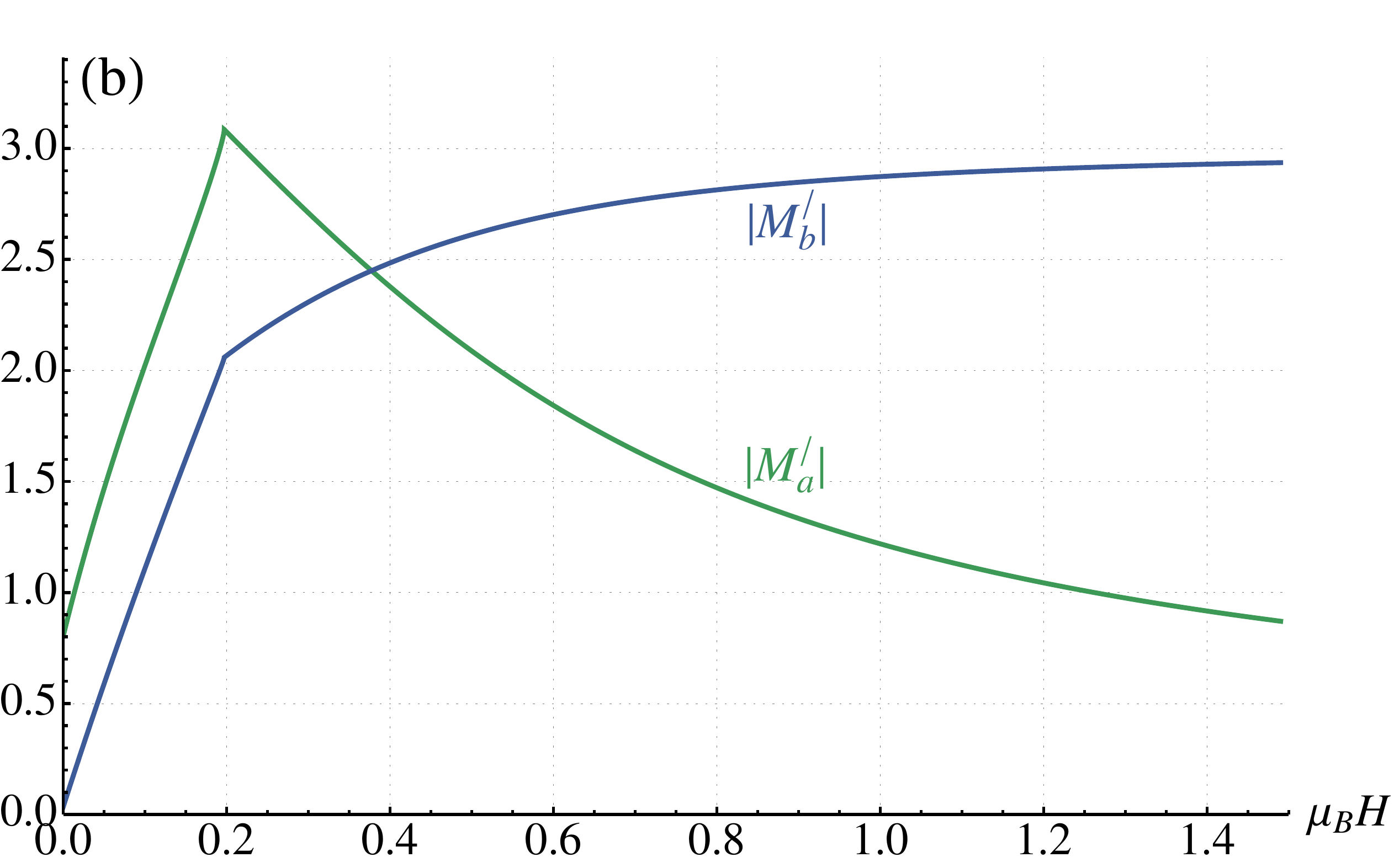}
\caption{Same as in Fig.~\ref{fig:SofQvsH} but now we show the behavior up to fields much higher than $H^\ast$. Panels (a) and (b) correspond again to the parameter points $A$ and $B$, respectively, of Fig.~\ref{fig:PT}\,(a).}
\vspace*{-0.25cm}
\label{fig:SofQvsH_WholeRegion}
\end{figure}

\vspace*{-0.3cm}
We now turn to the important question of the dependence of the characteristic field $H^\ast$ on the coupling parameters. To address this question we have calculated $H^\ast$ for various paths and the results are shown in Fig.~\ref{fig:HcFixed}. Panel (a) shows the evolution of $H^\ast$ with $K$, for fixed $\Gamma\!=\!-1$ and various values of $J$, while panel (b) shows the evolution of $H^\ast$ with $\Gamma$ for fixed $K\!=\!-1$ and the same set of $J$ values as in panel (a). 
The results from panels (a) and (b) can be summarized as follows. 
(i) $H^\ast$ tends to decrease with increasing $|K|$ and $|\Gamma|$. 
(ii) The rate of this decrease is almost zero at small $J$ and then increases with increasing $J$.  
(iii) $H^\ast$ has a much stronger dependence on $J$ compared to its dependence on $K$ and $\Gamma$. In particular, $H^\ast$ grows with increasing $J$, with an almost constant rate. 

These features can also be seen in panel (c), which shows the data collected from (a) and (b), with $J$ on the horizontal axis. 
This figure shows explicitly that for small enough $J$, $H^\ast$ is essentially independent of $K$ and $\Gamma$ [point (ii) above] and grows linearly with $J$ [point (iii) above]. 
It also shows that the deviation from this linear growth becomes larger for larger $J$, where a dependence on $K$ and $\Gamma$ shows up.

\vspace*{-0.25cm}
\section{Discussion}
\vspace*{-0.25cm}
The results presented here provide a consistent interpretation of the recent scattering experiments by Ruiz {\it et al}~\cite{Ruiz2017}. 
First, the analysis confirms the linear growth of a uniform zigzag component along ${\bf a}$, the rapid decline of the IC order, and the intensity sum rule. 
The latter two observations are signatures of the smallness of $J$ compared to $K$ and $\Gamma$.

Second, the significant growth of the zigzag component under the field does {\it not} stem from their linear coupling, as the zigzag amplitude at $H^\ast$ remains almost the same in the absence of this coupling (i.e., for $g_{ab}\!=\!0$). 
This shows a strong intertwining of the zigzag component to both the IC counter-rotating component and the longitudinal magnetization, which is the one most directly driven by the field.
Moreover, the zigzag component is already present at zero field (the same is true for the magnetization along ${\bf b}$) but, for weak $J$, its amplitude is too weak to be observed, consistent with Ref.~[\onlinecite{Biffin2014a}].

Third, the field acts effectively as a FM Heisenberg coupling that counteracts the effect of $J$, and the state reached at $H^\ast$ is qualitatively the same as the state stabilized by a negative $J$ at zero field.  An immediate ramification is that $H^\ast$ grows linearly with $J$, which is demonstrated numerically in the relevant regime of interest. 
Accordingly, the rapid decline of the IC component is another  signature of the smallness of $J$. 
In particular, the curve shown in Fig.~\ref{fig:HcFixed}\,(c) gives $J\!\simeq\!4$\,K for the experimental value of $H^\ast\!=\!2.8$\,T~\cite{Ruiz2017} (assuming $g_{bb}\!=\!2$ and $g_{ab}\!=\!0.1$; the latter does not affect $H^\ast$ as strongly as $g_{bb}$), which is indeed much smaller~\footnote{We should note that these {\it ab initio} calculations give a negative $J$, which on the basis of the $J$-$K$-$\Gamma$ model cannot deliver the IC order reported experimentally.} than the reported {\it ab initio} values of $K$ and $\Gamma$~\cite{Katukuri2016,KimKimKee2016, Tsirlin2018}.
%

Given this, the recent report~\cite{Tsirlin2018} that $H^\ast$ decreases with pressure is direct evidence (within the $J$-$K$-$\Gamma$ framework) that $J$ decreases under pressure. 
However, the behavior of $H^\ast$ alone cannot tell us what happens to $K$ and $\Gamma$, because $H^\ast$ shows a weak dependence on these couplings only at larger $J$; There is however independent evidence from $\mu$SR~\cite{Tsirlin2018} and {\it ab initio} calculations~\cite{KimKimKee2016,Tsirlin2018} showing that pressure increases the ratio $|\Gamma/K|$, with the system approaching the correlated classical spin liquid regime of the large-$\Gamma$ model~\cite{IoannisGamma}.

An experimental result that is not readily captured by the present classical description is the reported value of the magnetization at $H^\ast$ per Ir site, $m/\mu_B\!\simeq\!0.31$~\cite{Ruiz2017}, which is about two times smaller than the value $z_1(H^\ast)g_{bb} S\!\simeq\!\frac{1}{\sqrt{3}}$ (for weak $J$ and $g_{ab}\!=\!2$), deduced from the above analysis. 
Given that the IC component disappears at $H^\ast$, the discommensurations that are missed by the $K$-state ansatz at zero field~\cite{Sam2018} are not present any longer. 
Therefore, the above large disagreement in the magnetic moment $m$ should be attributed to the reduction of the spin length by quantum fluctuations. 
Such an unusually~\footnote{We know from other anisotropic models of this type (see e.g. Ref.~[\onlinecite{IoannisK1K2}]) that the spin length reduction by quantum fluctuations is otherwise very small due to the anisotropy spin gap.} large reduction can only be expected close to the frustrated region $\phi\!\sim\!3\pi/2$, which is in line with the remaining evidence above. A standard $1/S$ expansion around the $K$-state at $H^\ast$ should confirm this picture, but this is out of the scope of the present study.  

The broader emerging picture, taken together with the evidence from the zero-field study of Ref.~[\onlinecite{Sam2018}], gives strong confidence to the intuitive hypothesis made in \cite{Sam2018}, that the actual zero-field IC phase of $\beta$-Li$_2$IrO$_3$ is a long-wavelength twisting of a nearby commensurate state (the $K$-state). 
This route can be useful in analyzing related materials with analogous IC orders, such as $\alpha$-Li$_2$IrO$_3$ and $\gamma$-Li$_2$IrO$_3$~\cite{Singh2012,Biffin2014b,Modic2014,Williams2016,Satoshi2016}, and thus shed light to the delicate interplay between the various microscopic interactions, and map out the most relevant instabilities and their distinctive experimental fingerprints.


\vspace*{-0.3cm}
\section{Acknowledgments}
\vspace*{-0.3cm}
We are grateful to Samuel Ducatman for collaboration on related work, and to Cristian Batista, Radu Coldea and Alexander Tsirlin for valuable discussions. 
This work was supported by the U.S. Department of Energy,  Office of Science, Basic Energy Sciences under Award \# DE-SC0018056.

\bibliography{refs}\end{document}